\documentclass[]{JHEP3}
\usepackage{graphicx}
\usepackage{cite}
\title{Electroweak corrections and Bloch-Nordsieck violations in 2-to-2 processes at the LHC}
\author{W.J.~Stirling, E.~Vryonidou\\Cavendish Laboratory, J.J. Thomson Avenue,
Cambridge CB3 0HE, UK}
\preprint{Cavendish-HEP-2012/19}
\abstract
{We consider the effect of next-to-leading order (NLO) electroweak corrections to Standard Model $2\to 2$ processes, taking
into account the potentially large double logarithms originating from both real and virtual corrections. A
study of the leading Sudakov logarithms is presented and Bloch-Nordsieck (BN) violations are
discussed for processes at the CERN Large Hadron Collider. In particular, we focus on the processes $Z/\gamma+$jet 
and also the ratio of $Z$ to $\gamma$ production. This ratio is known to be insensitive to NLO QCD corrections
but this is not expected to be the case for the electroweak corrections. We also comment on 
the effect of
 electroweak corrections and the presence of BN violation for QCD processes, in particular dijet production, and also for
purely electroweak processes such as $W+H$ and $W+Z$ associated production.}
\maketitle

\begin{document}
\section{Introduction}
With the LHC in full operation, the experiments are collecting enormous amounts of data that are then 
used to obtain precision measurements of various observables. 
These measurements are used to search for New Physics and also to test the Standard Model (SM). 
In order to maximize the information that can be extracted in this way,
the uncertainty in the theoretical predictions needs to be reduced. This has motivated the calculation of many
 processes at NLO in perturbative QCD and in some cases at NNLO QCD. In addition to QCD corrections, 
electroweak corrections also need to be taken into account. 
At centre-of-mass energies of a few TeV these can become numerically important, since in many cases they are known to 
take the form of Sudakov-like logarithms $\sim \alpha_w\textrm{log}^2(\mu^2/M_W^2)$, 
with $\mu$ a typical hard scale in the process. This has been known for a long time, and consequently electroweak corrections have been 
considered for a wide range of hard-scattering processes.  

Virtual electroweak (EW) corrections have been calculated for a range of
SM processes such as $Z/\gamma+$jet
\cite{Kuhn:2004em,Kuhn:2005az,Kuhn:2005gv}, inclusive jet production
\cite{Moretti:2006ea}, top pair production \cite{Moretti:2006nf,Kuhn:2005it}, and
b$-$jet production \cite{Kuhn:2009nf}. Previous work has generally focused on the virtual EW
corrections. Real corrections are usually not taken into account, even though they contribute at the same order in 
perturbation theory. This approach can be justified by arguing that real emission of $W$
and $Z$ bosons, which subsequently decay, leads to qualitatively different final states with different (and more 
complex) signatures in the detectors. Moreover as the
$W$ and $Z$ masses act as infrared regulators, EW virtual corrections remain finite,
in contrast to QCD corrections where the addition of real corrections is necessary
to cancel the soft and collinear divergences. However, exclusion of the real emission contribution is only 
justified when {\it exclusive} final states are considered. In contrast, when fully
inclusive final states are studied, the effect of virtual corrections is
expected to be at least partially compensated by the effect of real electroweak gauge boson
emission. Nevertheless, and as we shall discuss in detail below, the cancellation is not expected to be exact due to violation
of the Bloch-Nordsieck (BN) theorem even for fully inclusive observables, see for example 
\cite{Ciafaloni:2000df,Ciafaloni:2000rp,Ciafaloni:2000gm,Ciafaloni:2001vt}. The BN
violation originates from the non-abelian nature of the electroweak charges
in the initial partonic state and is related to electroweak symmetry breaking. The same
violation is expected in QCD, as QCD is also a non-abelian theory. However in this case it has no physical consequences
because of the colour averaging of the initial partonic states. In the
electroweak case the initial-state weak isospin charge is fixed and there is no
averaging, which leads to non-cancelling Sudakov logarithms of the form $\sim \alpha_{w} \textrm{log}^2\frac{s}{M_V^2}$, 
with $V=W,Z$. Averaging over different possible weak isospin initial states (e.g.
$\sigma_{e\bar{e}}+\sigma_{\nu\bar{e}}$ or $\sigma_{ug}+\sigma_{dg}$) does lead to a full cancellation of the
BN violating Sudakov logs. In practice, for hadron colliders this averaging is not possible as 
different weak isospin (quark and gluon) states receive different parton distribution function (PDF) weights. 
As discussed in \cite{Ciafaloni:2000rp}, BN violating terms only originate from initial state radiation, with final state
radiation logarithms cancelling between virtual and real corrections. Moreover,
the BN violating logs are related only to $W$ boson emission. Sudakov logs 
from $\gamma/Z$ emission cancel between real and virtual corrections, as $\gamma/Z$ emission does not
change the particle type. 

The combined effect of real and one-loop virtual corrections for $Z$ and
$\gamma$ production, as well as for other SM processes such as inclusive jet and Drell-Yan production, has been considered in
\cite{Baur:2006sn}, where a partial cancellation is found. The effect of
selection cuts on the quantitative importance of BN violating Sudakov logs has been
considered in \cite{Bell:2010gi}. Even though the analysis focuses on the simpler
case of an $e^-e^+$ collider, the general conclusion drawn is that the extent of the compensation between the 
large (negative) logs of the virtual corrections and the large (positive) logs associated with the real
emission depends strongly on the phase space cuts applied to the real radiation, i.e. how much of the
collinear and soft $W$ and $Z$ radiation is allowed to escape detection.

In this paper we study how the real emission of soft EW bosons compensates the
large negative Double Logarithms (DL) associated with the virtual corrections in hard scattering
processes at the LHC. In Section~2 we review the emergence of Sudakov logs in virtual and real corrections 
and the origin of BN violations. 
In Section~3 we obtain the real corrections for
$\gamma/Z+1$~jet production at the LHC and combine them with the virtual logs. In Section~4, we comment on BN 
violation in other  $2\to 2$ scattering processes such as QCD dijet production and pure electroweak processes, 
before we conclude in Section~5. 

\section{Sudakov Double Logs in real and virtual corrections}

In this section we give a brief overview of Sudakov logs in EW corrections. In
order to do this at the DL level we need a recipe for the extraction of the DL
for both virtual and real corrections. For the real corrections this is
achieved by inserting the eikonal current for the emission of a soft gauge
boson of momentum $k$: 
\begin{equation}
J^{\mu}(k)=g_w\sum_{i=1}^n T_i \frac{p_i^{\mu}}{p_i k},
\end{equation} with $T_i$ the relevant isospin generator and the sum running over all external legs. Squaring the
 eikonal current and integrating over the phase space of the emitted gauge boson, 
we obtain a factor multiplying the LO result:
\begin{equation}
\int_M^E\frac{d^3
k}{2\omega_k(2\pi)^3}\frac{2p_ip_j}{(kp_i)(kp_j)}=\frac{1}{8\pi^2}\textrm{log}
^2\frac{2p_ip_j}{M^2},
\label{eikonal}
\end{equation}
in the high-energy limit of $M^2\ll 2 p_ip_j \sim E^2$, leading to a DL of the form
\begin{equation}
I(k)=\frac{\alpha_w}{4\pi}\sum_{i < j}T_i \cdot T_j\, 
\textrm{log}^2\frac{2p_ip_j}{M^2},
\label{realkey}
\end{equation} with $i$ and $j$ external legs and $\alpha_w=g_w^2/4\pi$ with the contribution being 
non-vanishing only when $T_i \cdot T_j$ is non-zero. In a Feynman diagram interpretation, for the real
corrections it is the interference between two diagrams where the gauge boson is emitted from two 
different external legs that leads to the DL. In practice, the correction takes the schematic 
factorised form: $\frac{\alpha_w}{4\pi} \textrm{L}^2 |M_{LO}|^2$ with $\textrm{L} = \textrm{log}^2(2p_ip_j/M^2)$, and an
appropriate prefactor determined by the couplings.

For the virtual corrections a generalised method for extracting soft and
collinear logs in the high-energy limit was presented in \cite{Denner:2000jv}. The DL originate from the
exchange of soft gauge bosons only between external legs. The method can be used for
any process and the change in the matrix element due to the loop correction is given by
\begin{equation}
\delta M^{i_1\dots i_n}=\frac{1}{2}\sum_{k=1}^{n}\sum_{l\ne
k}\sum_{V_a=A,Z,W^{\pm}} I^{V_a}_{i'_{k}i_k}(k)I^{\bar{V}_a}_{i'_l i_l}(l)
M_0^{i_1\dots i'_k\dots i'_l \dots
i_n}\frac{\alpha_w}{4\pi}\textrm{log}^2\frac{2p_kp_l}{M_{V_a}^2} ,
\label{virkey}
\end{equation}
for a process with $n$ external legs and tree-level LO matrix element $M_0^{i_1\dots i_n}$,
and $I^{V_a}_{i'_{k}i_k}(k)$ the SU(2)$\times$U(1) generators with $ieI^{V_a}_{i'_{k}i_k}(k)$ 
giving the coupling for the $V_a\bar{\phi}_i\phi_i'$ vertex, e.g. $I^{\gamma}_{e^+e^-}=1$. This expression
can be further simplified under certain assumptions leading to a single sum over
the external legs as shown in \cite{Denner:2000jv}:
\begin{equation}
\delta M^{i_1\dots i_n}=\sum_{k=1}^n -\frac{1}{2} 
C^{ew}_{i'_k i_k} M_0^{i_1\dots i'_k\dots i_n}\frac{\alpha_w}{4\pi}\textrm{log}^2\frac{s}{M_{W}^2},
\end{equation}
with $C^{ew}$ the Casimir operator as defined in \cite{Denner:2000jv} and where 
all logarithms of pure electromagnetic origin or of the form $\textrm{log}(M^2_Z/M_W^2)$ are neglected.
 In practice, we keep the first 
form given in Eq.~(\ref{virkey}) to facilitate the comparison with the real corrections. The
generator prefactors match between the corresponding combinations of external
legs in real and virtual corrections. For a fixed
initial state the extent of the BN violation  depends on which real emission processes are allowed, 
as $W^{\pm}$ emission might be forbidden due to charge conservation.
 When both emission from external leg $k$-absorption from leg $l$ (virtual) and
interference between the two emissions (real) are allowed, it is the difference between
$M_0^{i_1\dots i'_k\dots i'_l \dots i_n}M_0^{i_1\dots i_k\dots i_l \dots i_n}$
and $M_0^{i_1\dots i'_k\dots i_l \dots i_n}M_0^{i_1\dots i_k\dots i'_l \dots
i_n}$ that determines whether the BN theorem is violated. In the case of neutral gauge bosons ($\gamma,Z$), 
the two expressions are always the same as the particle type does not change, i.e. 
$i'_k=i_k$. This explains more formally the observation that only $W$ boson
emission leads to BN violating logs. 

Using Eq.~(\ref{virkey}) we can identify the
origin of the DL in the virtual corrections discussed in the following sections, with the prefactors obtained from the
 corresponding couplings between the gauge bosons and the emitting external particles. The
interference of $\delta M^{i_1\dots i_n}$ with the LO diagram gives the
contribution of the virtual corrections to the matrix element squared:
\begin{equation}
\delta |M^{i_1\dots i_n}|^2=\sum_{k=1}^{n}\sum_{l\ne k}\sum_{V_a=A,Z,W^{\pm}}
I^{V_a}_{i'_{k}i_k}(k)I^{\bar{V}_a}_{i'_l i_l}(l) M_0^{i_1\dots i'_k\dots i'_l
\dots i_n} M_0^{*i_1\dots
i_n}\frac{\alpha_w}{4\pi}\textrm{log}^2\frac{2p_kp_l}{M_{V_a}^2} ,
\label{virtot}
\end{equation} while the corresponding result for the real corrections takes the
slightly different form:
\begin{equation}
\delta |M^{i_1\dots i_n}|^2=-\sum_{k=1}^{n}\sum_{l\ne k}\sum_{V_a=A,Z,W^{\pm}}
I^{V_a}_{i'_{k}i_k}(k)I^{\bar{V}_a}_{i'_l i_l}(l) M_0^{i_1\dots i'_k\dots
i_n}M_0^{*i_1\dots i'_l\dots i_n}
\frac{\alpha_w}{4\pi}\textrm{log}^2\frac{2p_kp_l}{M_{V_a}^2}.
\label{reatot}
\end{equation}
These two expressions can be used to check whether a process will exhibit BN violations. 
We will use them for specific processes in the following sections.

\section{$\gamma/Z$ plus jet production}
\subsection{Partonic results}
We start our study by focusing on $\gamma/Z+1$~jet production, for which 
analytic results in the high-energy limit for the virtual corrections are already available 
in the literature~\cite{Kuhn:2004em,Kuhn:2005az,Kuhn:2005gv}. We will combine these with 
our real correction calculations. In this calculation $\gamma$ and
$Z$ are produced on-shell and no decays are taken into account. Also no pure 
electromagnetic corrections will be taken into account. We start by 
considering $\gamma+1$~jet production as an example. The virtual correction diagrams are 
shown in \cite{Kuhn:2005gv}. As discussed there,
the virtual corrections in the high-energy limit ($s,t,u  \gg M_{Z/W}^2$) take the form
of Sudakov logarithms, $\textrm{log}(r/M_V^2)$ where $M_V$ is the mass of the
exchanged boson and $r$ one of the Mandelstam variables $s,t,u$, following Eq.~(\ref{virkey}). 
The virtual corrections are extracted in the
approximation $M_Z=M_W$. To obtain the
corresponding leading logs from the real corrections, we work in the same limit with the 
eikonal current. For fixed initial state $q\bar{q}$ the real boson
emitted is always a $Z$, to conserve charge, while for the $qg$ initial state the emitted 
electroweak boson can be
either a $Z$ or $W$, with $W$ changing the flavour of the final state quark. 
In the following subsections for illustration purposes we present some analytic partonic results for a
selection of initial states, to examine the BN violating effects, while hadronic
numerical results will be presented in a subsequent subsection.

\subsubsection{Initial state: $q\bar{q}$}
We start by considering as an example $q(p_1)\bar{q}(p_2)\rightarrow \gamma(p_3) g(p_4)$. The
one-loop virtual corrections are calculated in \cite{Kuhn:2005gv} to
give the matrix element squared:
\begin{eqnarray}
\overline{\sum}|M^{q\bar{q}}|^2=8\pi^2\alpha_w\alpha_s
(N_c^2-1)\frac{\hat{t}^2+\hat{u}^2}{\hat{t}\hat{u}}[2Q_q^2+\frac{\alpha_w}{2\pi}A^
{(1)}]
\label{virtual}
\end{eqnarray}
with $A^{(1)}$ given in the high-energy limit by 
\begin{eqnarray}
A^{(1)}=-\sum_{\lambda=L,R}
Q_q[Q_q(C_{q\lambda}^{ew}-Q^2_q)(L_s^2-3L_s)+\frac{1}{s_w^2}T^3_{q\lambda}
(L_t^2+L^2_u-L_s^2)]
\end{eqnarray}with $s=2p_1p_2$, $t=-2p_1p_3$, $u=-2p_2p_3$, $L_r^k=\textrm{log}^k(|r|/M_W^2)$, $Q_q$ the charge,
$T^3_q$ the isospin generator, $C^{ew}$ the Casimir operator and $s_w=\textrm{sin}\theta_w$. 
The second term in $A^{(1)}$ comes from the exchange of a
virtual $W$ boson and is therefore non-zero only for left-handed quarks, while the
first is related to both $W$ and $Z$ exchange. The two contributions can be disentangled as
$(C_{q\lambda}^{ew}-Q^2_q)=\sum\limits_{Z,W^{\pm}}(I_V
I_{\bar{V}})_{q_{\lambda}}$. As an example, we show the $A^{(1)}$ results for $u\bar{u}$, 
substituting the corresponding couplings: 
\begin{equation}
A^{(1)}=-\frac{2}{3}\left[\frac{2}{3}\left(\left(\left(\frac{c_w}{2s_w}-\frac{
s_w}{6c_w}\right)^2+\frac{1}{2s_w^2}\right)+\frac{4s_w^2}{9c_w^2}
\right)(L_s^2-3L_s)+\frac{1}{2s_w^2}(L_t^2+L^2_u-L_s^2)\right],
\label{A1}
\end{equation}with $c_w=\textrm{cos}\theta_w$.
The virtual correction can be written as
\begin{equation}
 \delta^v \overline{\sum}|M^{u\bar{u}}|^2=-\frac{\alpha_w}{144\pi c_w^2s_w^2}\left[(9-24s_w^2+32s_w^4)L_s^2-9c_w^2L_s^2+27c^2_w(L_t^2+L_u^2)\right]\overline{\sum}|M^{u\bar{u}}_{LO}|^2,
\end{equation}with the first bracket coming from virtual $Z$ exchange and the rest from $W$ exchange.

\begin{figure}[h]
\centering
\includegraphics[scale=0.5]{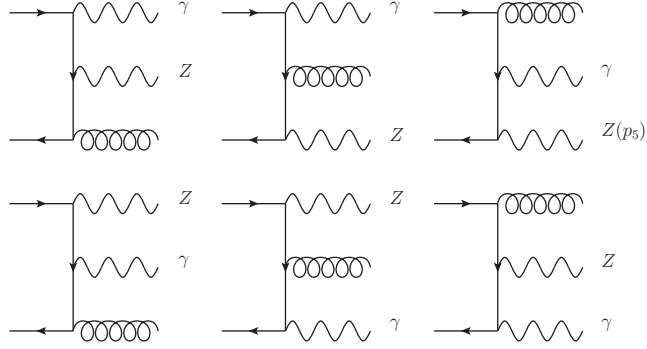} 
\caption{Feynman diagrams for $u(p_1)\bar{u}(p_2)\rightarrow A(p_3) g(p_4) Z(p_5)$.}
\label{Feyn2}
\end{figure}

The real correction contributions are obtained from the relevant Feynman diagrams. In
this case, because of charge conservation the only allowed emission is that of a
$Z$ boson as shown in Fig.~\ref{Feyn2}. 
Taking the appropriate limit of soft $p_5$, the $2\to 3$ correction to the matrix element squared factorises into
\begin{eqnarray}
\delta^r \overline{\sum}|M^{u\bar{u}}|^2&=&\overline{\sum}|M^{u\bar{u}}_{LO}|^2 \frac{\pi
\alpha_w}{18c_w^2s_w^2}(9-24s_w^2+32s_w^4) \frac{2p_1p_2}{(p_1p_5)(p_2p_5)}
\end{eqnarray}and integrating over $p_5$
\begin{eqnarray}
&\rightarrow &\frac{
\alpha_w}{144\pi c_w^2s_w^2}(9-24s_w^2+32s_w^4) L_s^2 \overline{\sum}|M^{u\bar{u}}_{LO}|^2  ,
\label{real}
\end{eqnarray}
with $p_1$ and $p_2$ the quark and antiquark momenta respectively.
Here we can identify the eikonal factor of Eq.~(\ref{eikonal}) which leads to a double
logarithm of the form $\textrm{log}^2(s/M_Z^2)$ when integrated over the
phase space of the soft $Z$. The $s$-log is expected from the Feynman diagrams,
as the emission of the soft $Z$ can occur from the both initial state legs and
the logarithm arises from the interference of the two corresponding diagrams. 
The polynomial in $\sin\theta_w$ prefactor, $9-24s_w^2+32s_w^4$, is determined by the $u-$quark couplings to the $Z$ and
 is in fact proportional to $c_A^2+c_V^2$. For the $d\bar{d}$ initial state this would change to $9-12s_w^2+8s_w^4$.
It is clear from  Eqs.~(\ref{A1}) and (\ref{real}) that the
cancellation between real and virtual corrections logarithms is not exact. Virtual
corrections include logarithms of all $s,t$ and $u$. By checking the DL term by
term, we can verify that the logarithms related to virtual soft $Z$ exchange -- the first and
third terms of the first bracket in Eq.~(\ref{A1}) --  are exactly cancelled by the
real corrections. The remaining terms are due to the exchange of virtual $W$ bosons and these cannot be
cancelled simply due to the absence of real $W$ emission diagrams. 

\subsubsection{Initial state: $qg$}
\begin{figure}[h]
\centering
\includegraphics[scale=0.5]{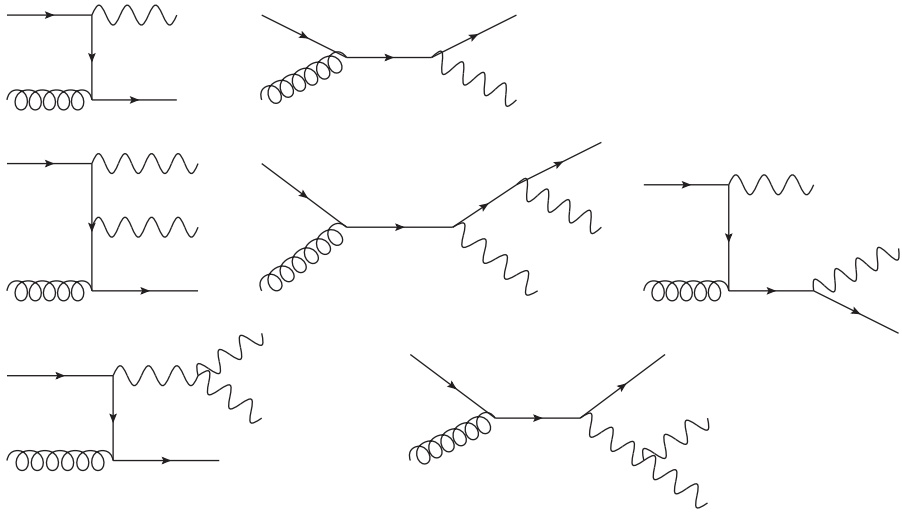} 
\caption{Feynman diagrams for $qg\rightarrow \gamma q$.}
\label{Feyn}
\end{figure}
 The topology of the relevant Feynman
diagrams for initial-state $qg$, which is found to be the dominant process for
$\gamma+1$~jet production at the LHC, is shown in Fig.~\ref{Feyn} with the LO diagrams in the first row and the
(NLO) real emission diagrams in the second and third rows, where an additional electroweak gauge boson is
emitted. The topology of the last two diagrams only applies for $W$ emission.
In the case of $ug\rightarrow \gamma u$ the real EW corrections include both
$ug\rightarrow \gamma u Z$ and $ug\rightarrow \gamma d W^+$. 
Calculating the matrix element for $
u(p_1)g(p_2)\rightarrow \gamma(p_3) u(p_4) Z(p_5)$ we obtain  
\begin{equation}
\delta^r \overline{\sum}|M^{ug}|^2=\overline{\sum}|M^{ug}_{LO}|^2 \frac{
\alpha_w}{144\pi c_w^2s_w^2}(9-24s_w^2+32s_w^4) \textrm{log}^2\frac{
2p_1p_3}{M_Z^2}.
\label{Zexc}
\end{equation} Note that the prefactor $(9-24s_w^2+32s_w^4)$ is the same as for the $u\bar{u}$ initial state.
The result can be obtained from Eq.~(\ref{real}) by appropriate use of crossing relations and
modification of the colour averaging factor at LO.  

For the $ u(p_1)g(p_2)\rightarrow d(p_3) \gamma(p_4) W^+(p5)$ process we obtain
 \begin{equation}
\delta^r \overline{\sum}|M^{ug}|^2=\overline{\sum}|M^{ug}_{LO}|^2 \frac{\alpha_w}{32s_w^2 \pi}[3
L_u^2-2
L_t^2+6
L_s^2], 
\label{Wexc1}
 \end{equation}in the soft $W$ limit. The prefactor of each of the logarithms is determined
by the couplings/charge of the quark involved. In a similar way results for all the
other processes, e.g. $u\bar{d}\rightarrow \gamma g W^+$, can be obtained by crossing relations. 
The results for the virtual corrections can be obtained from Eq.~(\ref{virtual}) by making the substitutions: 
$s\rightarrow u=-2p_1p_3, t\rightarrow s=2p_1p_2=2p_3p_4, u\rightarrow t=-2p_1p_4$ in the high-energy limit, 
giving for the leading-log (LL) result:
\begin{eqnarray}
\delta^v \overline{\sum}|M^{ug}|^2& = & -\frac{\alpha_w}{144\pi c_w^2s_w^2}\left[(9-24s_w^2+32s_w^4)L_u^2-9c_w^2
L_u^2+27c^2_w(L_s^2+L_t^2)\right]\overline{\sum}|M^{ug}_{LO}|^2 . \nonumber \\
& & 
\label{virtualug}
\end{eqnarray}

By comparing Eq.~(\ref{Zexc}) to Eq.~(\ref{virtualug}) we again see that the logarithms due to $Z$ exchange cancel, 
with the appropriate
decomposition to $W$ and $Z$ contributions. Furthermore the logarithms related to final state radiation, 
i.e. the term proportional to
$\textrm{log}^2(\frac{2p_3p_4}{M_W^2})=\textrm{log}^2(\frac{s}{M_W^2})$ in
Eq.~(\ref{Wexc1}), exactly cancels the equivalent term which appears in 
Eq.~(\ref{virtualug}). This is a good demonstration of the conclusions reached in \cite{Ciafaloni:2000rp},
regarding the difference between $W$ and $Z$ emission and also the final state radiation logs.

\subsubsection{Isospin averaging -- cancellation of double logarithms}
Another important check for the analytic calculations is the cancellation of the
logarithms on averaging over two states of the same weak isospin doublet. As an example, we take the
results for $ug\rightarrow \gamma u$ and $dg\rightarrow \gamma d$ with their
corresponding virtual and real corrections. Even though for the two processes
separately the sum of real and virtual corrections is BN violating, when we sum
the two with equal weights the logarithms cancel. 
This can be seen explicitly by considering the corresponding results for $ dg\rightarrow
\gamma d$:
\begin{equation}
\delta^v \overline{\sum}|M^{dg}|^2=-\frac{\alpha_w}{144\pi c_w^2s_w^2}
\left[(9-12s_w^2+8s_w^4)L_u^2-36c_w^2L_u^2+54c^2_w(L_s^2+L_t^2)\right]\overline{\sum}|M^{dg}_{LO}|^2,
\label{virtualdg}
\end{equation}
\begin{equation}
\delta^r \overline{\sum}|M^{dg}|^2=\overline{\sum}|M^{dg}_{LO}|^2 \frac{ \alpha_w}{144\pi
c_w^2s_w^2}(9-12s_w^2+8s_w^4) L_u^2 ,
\label{Zexcd}
\end{equation}
\begin{equation}
\delta^r \overline{\sum}|M^{dg}|^2=\overline{\sum}|M^{dg}_{LO}|^2 \frac{\alpha_w}{8s_w^2 \pi}[3
L_s^2-2
L_t^2+6
L_u^2] .
\label{Wexc2}
 \end{equation}

Adding up all relevant expressions for $ug$ (Eqs.~(\ref{Zexc}), (\ref{Wexc1}) and (\ref{virtualug}))  
and $dg$ (Eqs.~(\ref{virtualdg}), (\ref{Zexcd}) and (\ref{Wexc2})) 
and taking into account the factor of 4 difference between $|M^{dg}_{LO}|^2$ and $|M^{ug}_{LO}|^2$, due to electric charge, 
we find that no Sudakov logs remain uncancelled.
Of course in practice this averaging is not possible, as in hadronic collisions $u-$
and $d-$quark contributions are weighted by the corresponding $u-$ and $d$-quark PDFs 
which are of course different. The importance of the BN
violating terms is then determined by the values of the PDF momentum fractions and
factorisation scales that dictate the relative size of the PDF values, and therefore change the
relative cancellation between different flavours. 

\subsection{Hadronic results}
\begin{figure}[h]
\centering
\includegraphics[scale=0.6]{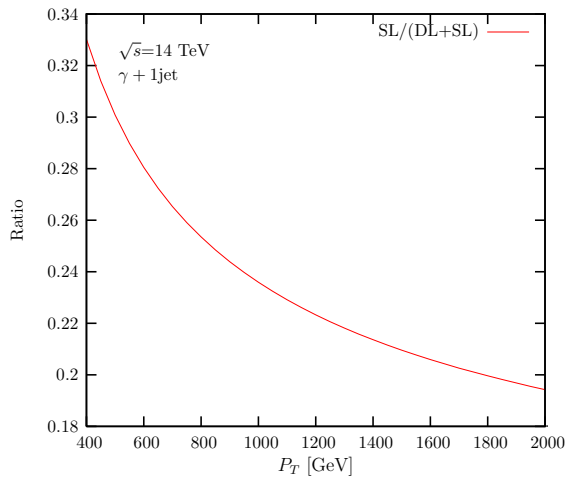} 
\caption{ Relative contribution of single logarithms to the virtual corrections in $\gamma+1$~jet production.}
\label{SL}
\end{figure}

$Z+1$~jet production can be treated in a similar way. Partonic results have been extracted and shown to lead to the same conclusions 
on BN violations, 
 but we do not show the expressions here as these are more lengthy due to the more complex structure of the $Z$ coupling to quarks.
The partonic results can then be used to obtain hadronic cross sections by convoluting with the appropriate PDFs.
For consistency we note that as we have not included any single logs (SL) from the
real corrections, we only keep the DL in the expressions for the virtual
corrections and therefore our results are valid only at the LL level. However we do need to keep in mind that the SL contribution remains
important even at very high $p_T$. This is shown in Fig.~\ref{SL}, where the fraction
of the virtual corrections to $\gamma+1$~jet production coming from the SL terms is shown for the LHC (at 14~TeV)  
as a function of the photon $p_T$, with no cuts applied. We see that the next-to-leading logarithms remain 
important even at high transverse momentum (at 2~TeV the contribution is still more than 20\%).

\begin{figure}[h]
\centering
\includegraphics[scale=0.6]{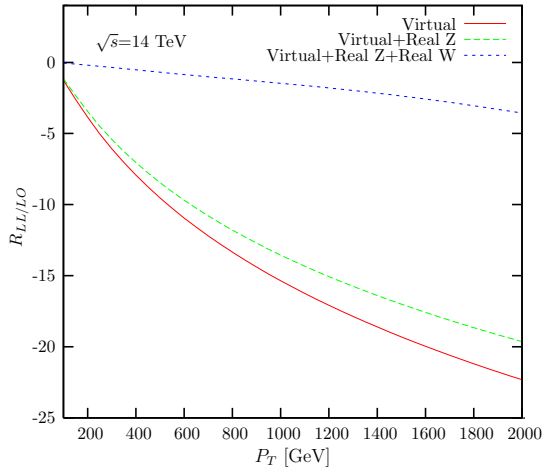} 
\caption{Virtual and real electroweak corrections to $\gamma+1$~jet production at 14~TeV.}
\label{all}
\end{figure}

The results for the $p_T^{\gamma}$ differential distribution for $\gamma+1$~jet production at the LHC, with only the virtual
corrections of \cite{Kuhn:2005gv} and then with the addition of the real corrections are shown in Fig.~\ref{all}, 
with $R_{LL/LO}=LL/LO-1$. These are
obtained using MSTW2008LO PDFs \cite{Martin:2009iq} and setting the factorisation and renormalisation scales 
to $p_T^{\gamma}$. We note that the virtual corrections decrease the cross
section by up to 20\% at very high $p_T$ (2~TeV) due to the large negative Sudakov
logs. We note that no phase space constraints are imposed on the emitted gauge bosons
other than the eikonal current approximation used to extract the leading logarithms. This computational 
set-up does not allow the application of cuts on the emitted gauge bosons or their decay products as we work 
in a two-body phase space approximation, with the LO result multiplied by a factor of the form $(1+ C \alpha_w \textrm{log}^2(r/M^2))$.
We note that adding the real correction due to $W$ emission has a bigger effect than that due to $Z$ emission. 
This is a combination of the different quark couplings to $W$ and $Z$ and the fact that including the $\mathcal{O}(\alpha_W^2)$ real $W$ 
corrections involves including new initial-state parton combinations e.g. $u\bar{d}$,
$d\bar{u}$ etc., which are not allowed at LO and these increase the cross section. The calculation is performed using
a diagonal CKM matrix. Evidently there is a significant but not exact compensation
 of the negative virtual contribution by the real corrections at the DL level.

\begin{figure}[h]
 \centering
\includegraphics[scale=0.6]{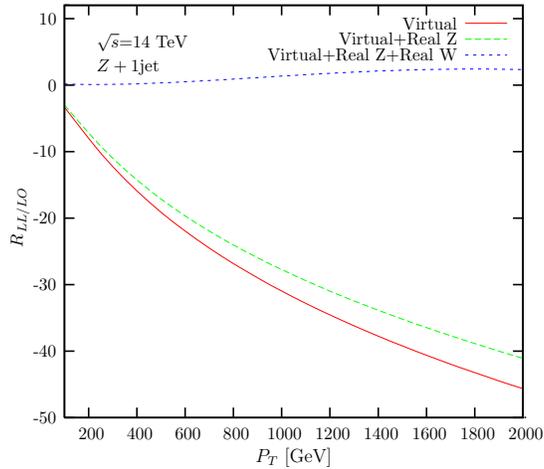} 
\caption{Virtual and real electroweak corrections to $Z+1$~jet production at 14~TeV.}
\label{allZdl}
\end{figure}

The same procedure is followed for $Z+1$~jet, with the analytic DL results of
the partonic cross sections convoluted with the PDFs. In the calculation
of the matrix elements for the partonic cross sections, we ignore the mass of
the $Z$. This is a good approximation in the region of high $p_T$ where the LL approximation is most valid. 
The results are shown in
Fig.~\ref{allZdl}. In this case it can be seen than the real corrections
{\it overcompensate} the negative double logarithms of the virtual corrections, with the $W^{\pm}$ contribution being dominant.

\subsubsection{Comparison with numerical calculation of full real corrections}
Our results for the real corrections have been obtained using the eikonal current to extract the leading logarithmic terms.
 This can be compared to the full real corrections result obtained using the exact matrix elements for the $2\to 3$ processes 
and therefore a 3-body phase space. This can be done using publicly available programmes such as MCFM~\cite{MCFM}. 
This combination of the real and virtual corrections has been first discussed in  \cite{Baur:2006sn} for a series of SM processes, 
including $\gamma/Z+$jet production. 

\begin{figure}[h]
\begin{minipage}[b]{0.5\linewidth}
\centering
\includegraphics[scale=0.6]{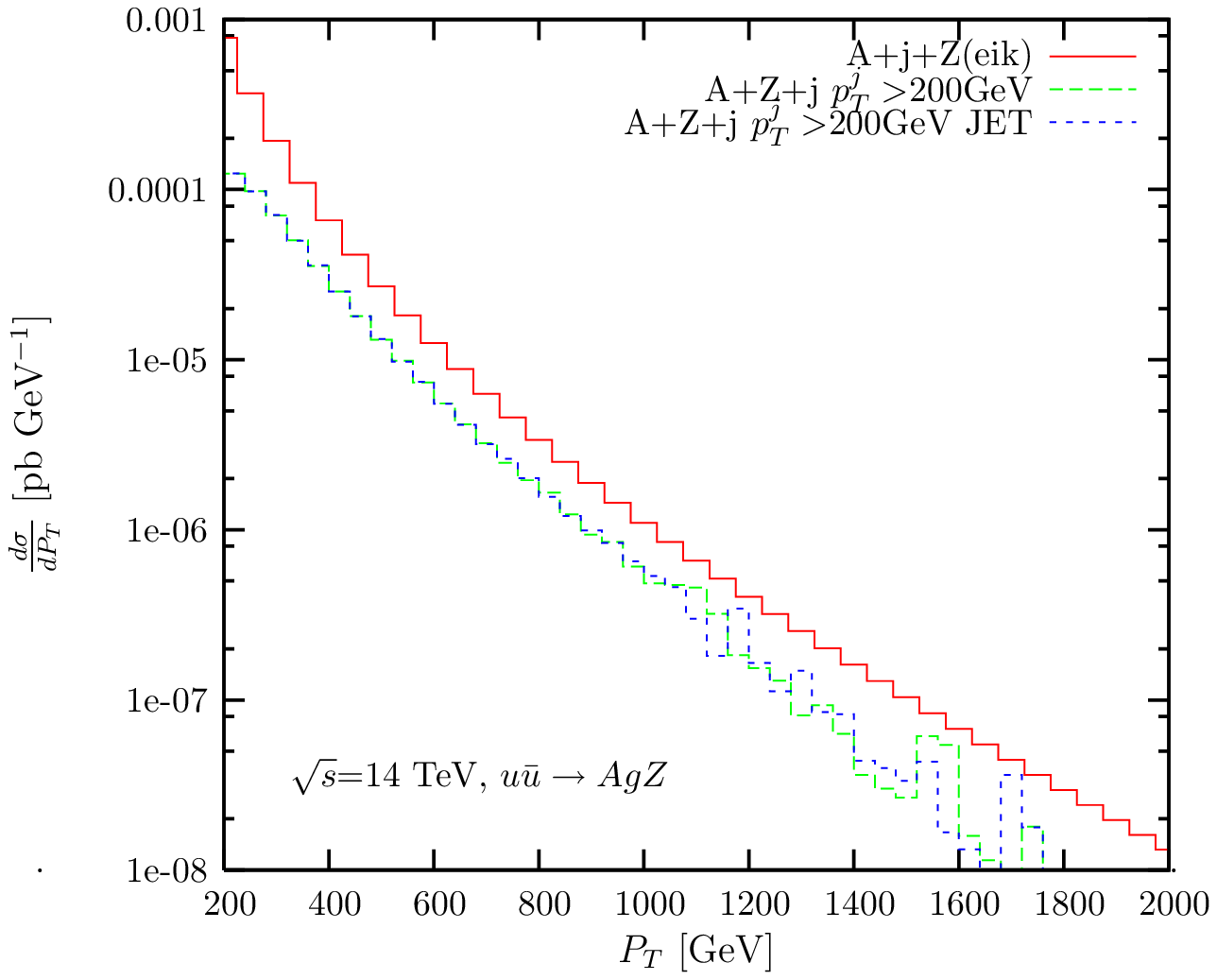} 
\caption{Photon transverse momentum distribution for $u\bar{u}\rightarrow AgZ$, for soft $Z$ (eikonal current) 
and using MCFM, with fixed scales. }
\label{uubAZ}
\end{minipage}
\hspace{0.5cm}
 \begin{minipage}[b]{0.5\linewidth}
 \centering
\includegraphics[scale=0.6]{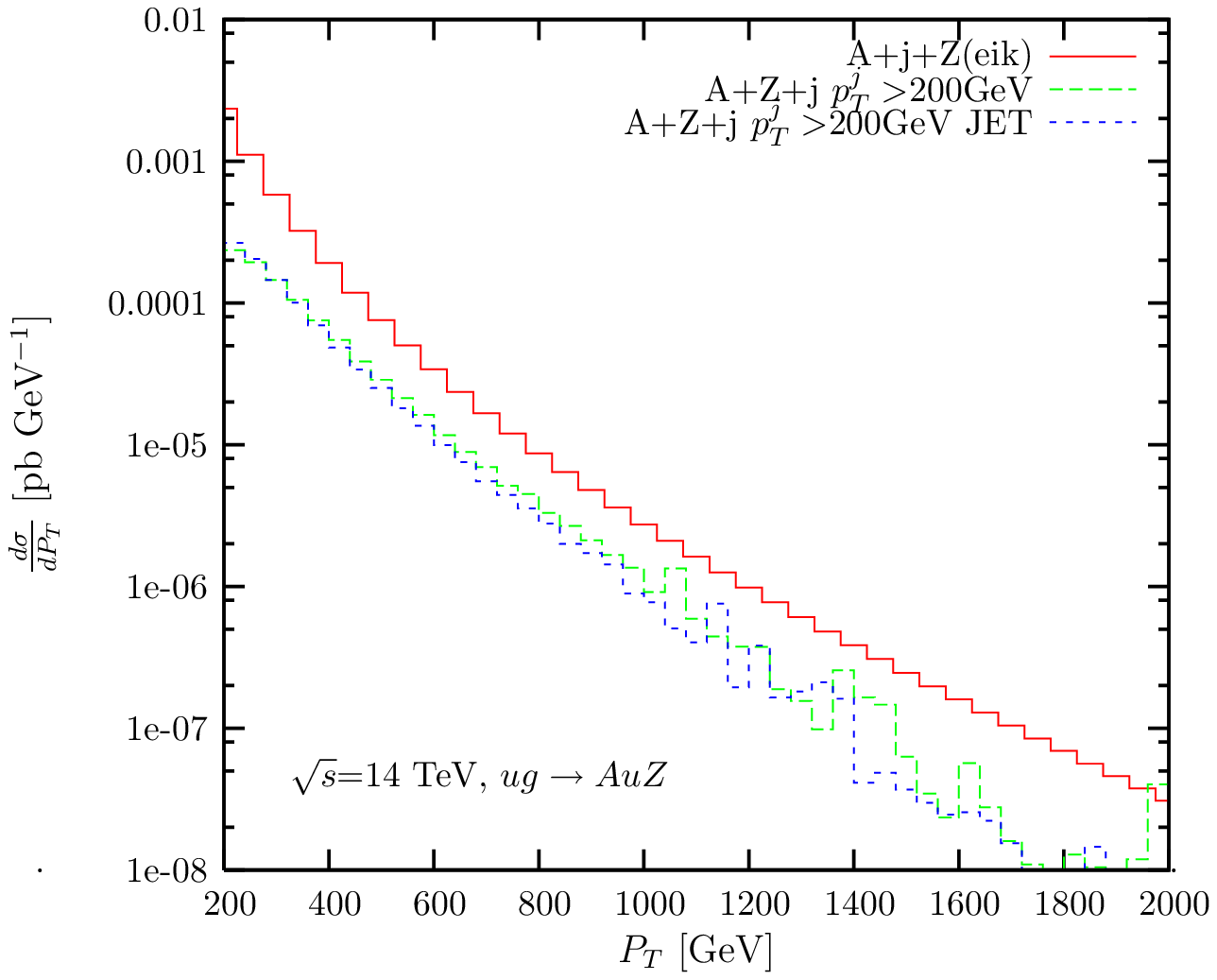} 
\caption{Photon transverse momentum distribution for $ug\rightarrow AuZ$, for soft $Z$ (eikonal current) and using MCFM, with fixed scales.}
\label{ugAZ}
\end{minipage}
\end{figure}

\begin{figure}[h]
\begin{minipage}[b]{0.5\linewidth}
\centering
\includegraphics[scale=0.6,trim=1cm 0 0 0]{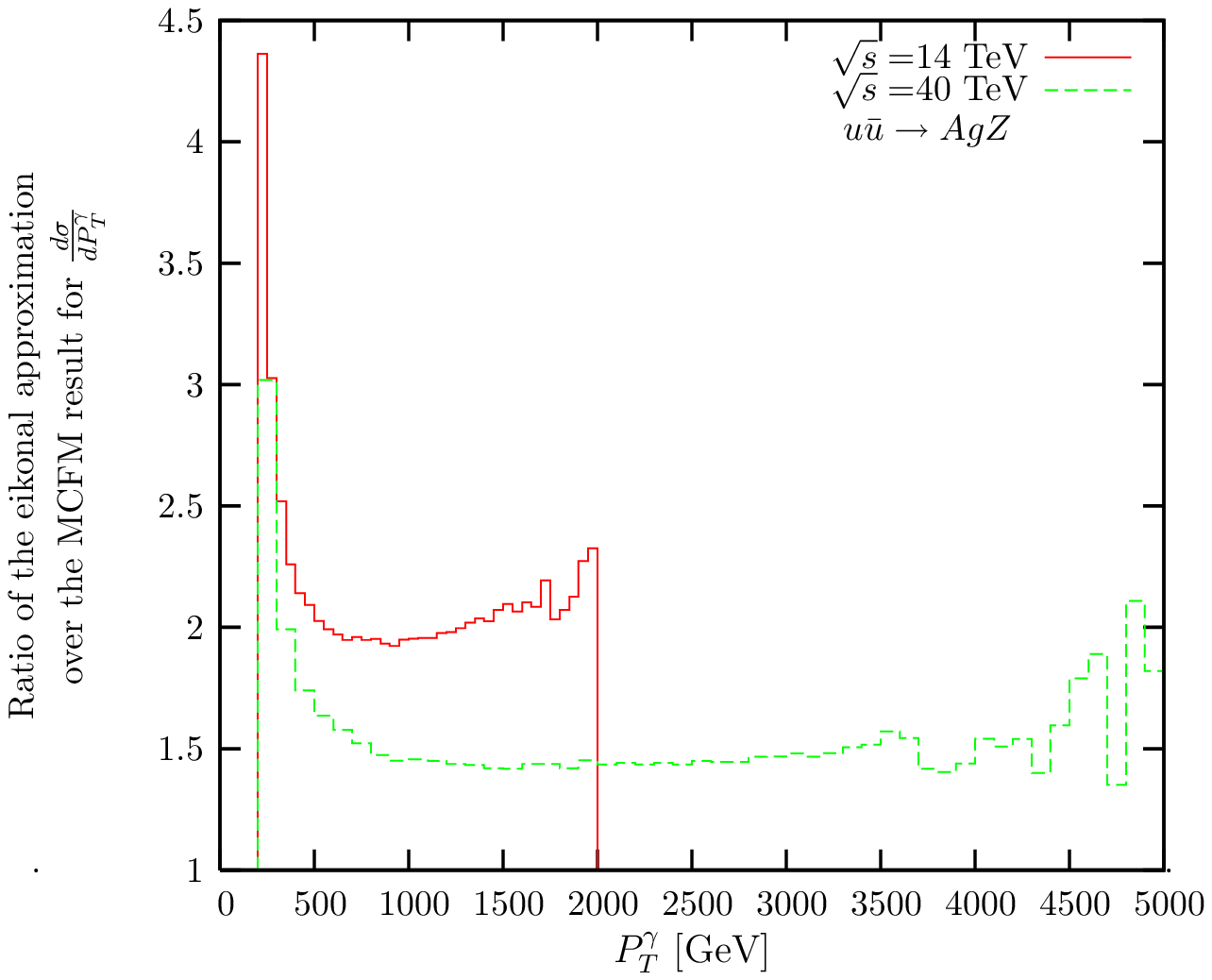} 
\caption{Ratio of the eikonal approximation result to the $2\to 3$ MCFM result for the subprocess $u\bar{u}\rightarrow AgZ$.}
\label{ratio}
\end{minipage}
\hspace{0.5cm}
 \begin{minipage}[b]{0.5\linewidth}
 \centering
\includegraphics[scale=0.6]{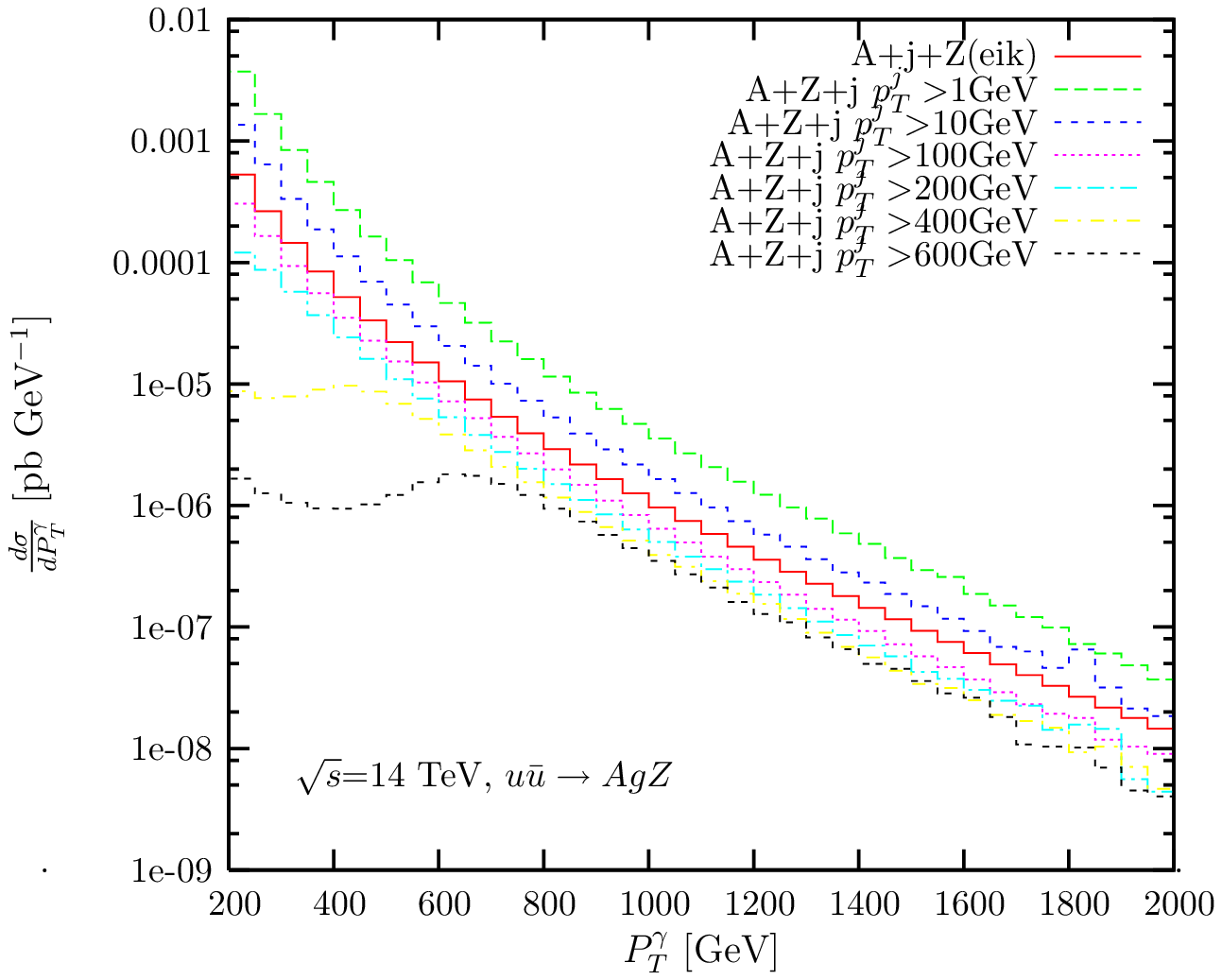} 
\caption{Effect of the jet $p_T$ cut on the photon transverse momentum distribution for subprocess $u\bar{u}\rightarrow AgZ$. }
\label{jetcut}
\end{minipage}
\end{figure}

Here we select two specific processes, corresponding to initial states $u\bar{u}$ and $ug$, 
to gauge the differences in the two approaches. We use the same PDFs, scale and EW parameters. 
The results are shown in Figs.~\ref{uubAZ} and \ref{ugAZ}, where we show the photon $p_T$ distribution 
for the eikonal result and the jet and photon $p_T$ distributions we obtain using MCFM and hence the 
full $2\to 3$ matrix elements, imposing a cut of 200~GeV on the jet $p_T$. The ratio of the two photon $p_T$ 
distributions is shown in Fig.~\ref{ratio} for centre-of-mass energies of 14~TeV and 40~TeV. At the lower collider energy we terminate 
the distribution at  $p_T \sim  2$~TeV as we run into low statistics. 
In this plot we also include the ratio for the corresponding process obtained for a centre-of-mass energy of 40~TeV. 
Evidently the two results agree much better at 40~TeV, as the eikonal condition is better satisfied at higher energies. 

Another effect we can study is the impact of the cut on the jet transverse momentum on the photon $p_T$ distribution. 
This is shown for $u\bar{u}\rightarrow AgZ$ in Fig.~\ref{jetcut}. We see that as expected increasing the cut decreases 
the cross section and also modifies the shape at low $p_T$.

\subsubsection{The $\gamma$ to $Z$ ratio}
\begin{figure}[h]
\centering
\includegraphics[scale=0.7]{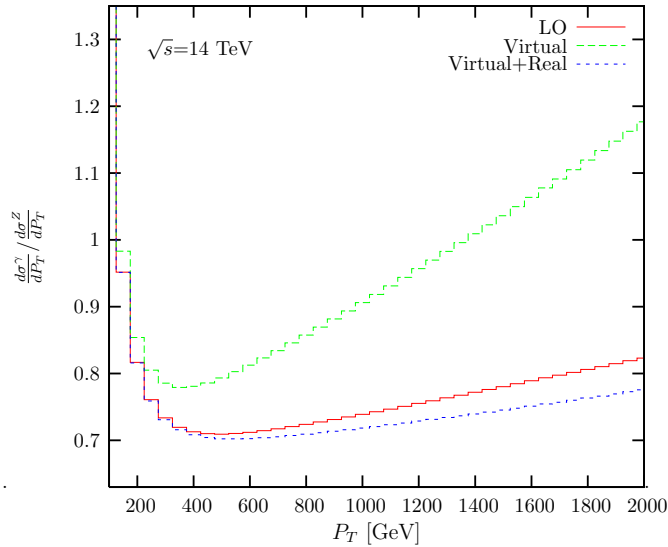} 
\caption{Ratio of $\gamma+$jet to $Z+$jet cross sections as a function of $p_T$ at LO, including both virtual corrections 
and real corrections at the DL level.}
\label{ratiogZ}
\end{figure}

The importance of the ratio of the $\gamma+$~jets to $Z+$~jets cross sections in new physics searches 
has been discussed in \cite{Ask:2011xf}. The ratio is useful to 
better calibrate the missing transverse energy plus jets background to SUSY production coming from $Z(\rightarrow$ neutrinos) plus jets.
Leading order results have been shown in \cite{Ask:2011xf} for the ratio of
$\gamma$ to $Z$ for one or more jets. The ratio is known to be insensitive to
NLO QCD corrections, and suffers less than the individual cross sections from scale and PDF uncertainties. 
The effect of virtual corrections on this ratio  has been considered in \cite{Kuhn:2005gv}.
The
results are shown in Fig.~\ref{ratiogZ}. We note that virtual corrections
increase the ratio at large $p_T$ because the $Z$ cross section is decreased
more than the $\gamma$ cross section by the virtual DL. However, the inclusion of soft real
radiation in the eikonal limit significantly shifts the result closer to the LO value. In practice, this of course
depends on the amount of soft and collinear radiation that is allowed to escape detection.
Setting cuts on the emitted bosons and their decay products reduces the effect of the real corrections
leading to a result closer to the virtual result of \cite{Kuhn:2005gv}. 
A detailed study including cuts and experimental restrictions, which is beyond the scope of this work, would be needed to 
determine how rapidly this convergence occurs.

\section{Bloch-Nordsieck violations in other hard-scattering processes}
\subsection{QCD dijet production}
Electroweak corrections for dijet production have been studied in \cite{Moretti:2006ea},\cite{Scharf:2009sp} and recently in \cite{Dittmaier:2012kx}. 
The virtual one-loop EW corrections ($\mathcal{O}(\alpha_s^2\alpha_w)$) to the single
jet inclusive cross section have first been studied in \cite{Moretti:2006ea}. These
are found to decrease the LO cross section
($\mathcal{O}(\alpha_s^2+\alpha_s\alpha_w+\alpha_w^2)$) by up to 35\% at very
large jet energies($\sim$4~TeV) at the LHC. It is noted in \cite{Moretti:2006ea}
that $W-$ and $Z-$bremsstrahlung effects might counterbalance the negative logs
of the virtual corrections, as discussed in \cite{Ciafaloni:2006qu}. In this case
the calculation of the $\mathcal{O}(\alpha_s^2\alpha_w)$ result is complicated
by the presence of the LO purely electroweak processes increasing the number of Feynman diagrams, which lead to different
possible interferences that must be taken into account. At tree level the contribution of $\mathcal{O}(\alpha_s\alpha_w+\alpha_w^2)$ terms is found in \cite{Moretti:2006ea} to reach 15\% of the QCD contribution at 4 TeV.

In this section we present a qualitative discussion focused on the different subprocesses contributing to
$2\to 2$ QCD scattering and the corresponding EW corrections at the DL level, examining
which subprocesses exhibit BN violations and their origin, based on a Feynman diagram approach.  However in order
to simplify the
discussion we ignore the pure LO EW contributions, so that the 
$\mathcal{O}(\alpha_s^2\alpha_w)$ corrections come from the interference of the
LO  $\mathcal{O}(\alpha_s)$ matrix element with the NLO
$\mathcal{O}(\alpha_s\alpha_w)$ matrix element. To identify which processes exhibit BN violations, we need to focus only on virtual and real
$W$ corrections as the $Z$ corrections always cancel, as shown by the
form of Eqs.~(\ref{virtot}) and (\ref{reatot}). Here we briefly consider
the subprocesses as organised in \cite{Moretti:2006ea}. The classification is based on 
the initial state parton combinations which determine the PDF weight. The
examples given are for the first generation of quarks, but the same arguments hold for the 
second generation. In what follows, the CKM matrix is taken to be diagonal. Quantitative results will be presented in another study.
 We note that consideration of the full real corrections involves calculating processes such as $W/Z+2$~jets production.
\begin{itemize}
 \item \underline{$gg\rightarrow gg$} This subprocess receives no real or virtual
corrections at $\mathcal{O}(\alpha_s^2\alpha_w)$.
\item \underline{$gg\rightarrow q\bar{q}$} The corrections consist of real and virtual
$W$ emission from the final-state legs. As we are summing over all quark
flavours in the final state the divergences cancel. This can be inferred from the
form of Eqs.~(\ref{virtot}) and (\ref{reatot}): both the real and virtual
corrections are proportional to $M_0^{ggd\bar{d}}M_0^{*ggu\bar{u}}$ and
therefore the cancellation of Sudakov logs is exact.
\item \underline{$q\bar{q}\rightarrow gg$} As the initial state is fixed to be a same
flavour quark-antiquark pair, real $W$ emission is not allowed in this subprocess.
Therefore the DL coming from the exchange of a soft $W$ boson between the initial state
quarks, as shown in Fig.~\ref{qqbarW}, remain uncancelled. We note that processes of type $u\bar{d}\rightarrow W^+ g g$ are considered separately below. 
\begin{figure}[h]
\centering
\includegraphics[scale=0.7]{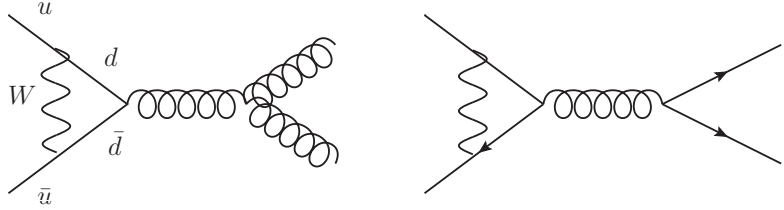} 
\caption{Virtual correction diagrams for jet production from initial state $u\bar{u}$.}
\label{qqbarW}
\end{figure}
\item \underline{$qg\rightarrow qg$ ($\bar{q}g\rightarrow \bar{q}g$)} Real emission takes place from the initial- and final-state quark legs. 
Real and virtual
corrections are both proportional to $M_0^{ugug}M_0^{*dgdg}$ leading to exact cancellation between
real and virtual corrections, and therefore no BN violation is expected.
\item  \underline{$qq\rightarrow qq$ ($\bar{q}\bar{q}\rightarrow \bar{q}\bar{q}$)} BN
violating logs are present. This can be seen by noting that for the real
corrections we obtain the interference between the diagram where $W$ is emitted
from $q(p_1)$ and that where $W$ is emitted from $q(p_2)$, leading to a term
proportional to $\textrm{log}^2(2p_1p_2/M_W^2)$. In the virtual corrections no Feynman
diagram can be drawn with $W$ exchanged between the initial state legs and
therefore the real DL remain uncancelled. The remaining  $t$ and $u$ double logs cancel. (This subprocess is in general complicated
by the presence of the two diagrams at LO ($u-$ and $t-$channel), with the real corrections not being
proportional to $|M_{LO}|^2$.) 
\item \underline{$qQ\rightarrow qQ$ ($\bar{q}\bar{Q}\rightarrow \bar{q}\bar{Q}$), with $q,
Q$ in the same generation}. In this case the vertex correction diagrams exactly
cancel the interference of diagrams with emission from the same flavour fermion
line. However as we have quarks of the same generation a set of box diagrams is
also allowed. These are shown in Fig.~\ref{boxes}. There is no corresponding set
of real emission diagrams to cancel the divergences from these diagrams, as a $W^+$ is emitted from one flavour leg but a $W^-$ from the other,
 and we therefore expect BN violating logs. 
\begin{figure}[h]
\centering
\includegraphics[scale=0.7]{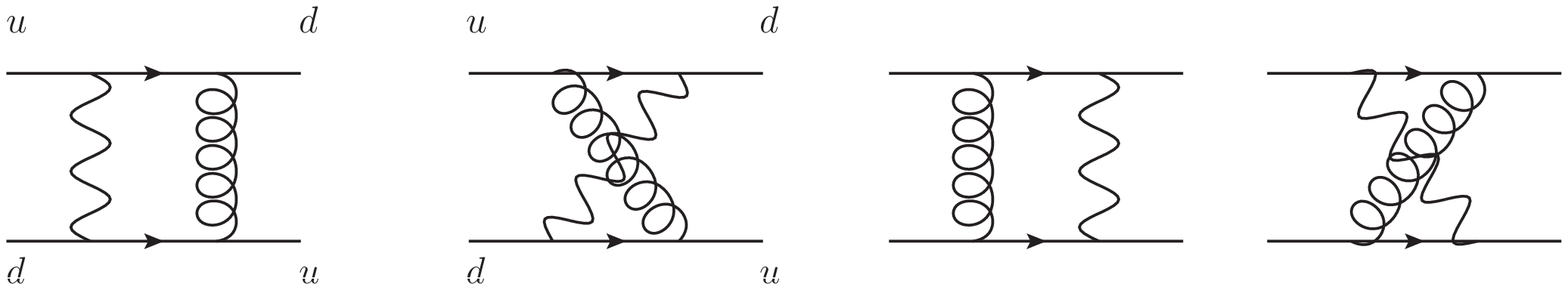} 
\caption{Box diagram corrections for $qQ\rightarrow qQ$ scattering.}
\label{boxes}
\end{figure}
\item \underline{$qQ\rightarrow qQ$ ($\bar{q}\bar{Q}\rightarrow \bar{q}\bar{Q}$) different
generation} Only the $t-$channel gluon exchange diagram exists at LO. The virtual corrections
consist only of the vertex corrections diagrams with the divergences exactly
cancelling those of the real emission diagrams. Interference occurs only between the diagrams where 
emission comes from the same flavour initial and final legs. No BN violation is therefore 
expected from this subprocess.
\item \underline{$q\bar{Q}\rightarrow q\bar{Q}$ different generation} The same conclusions as in
the previous subprocess apply, and no BN violation is expected.
\item \underline{$q\bar{Q}\rightarrow q\bar{Q}$ same generation} In this case we note that
for initial state $q\bar{Q}$ at
$\mathcal{O}(\alpha_s^2\alpha_w)$ the real corrections involve processes of type
$u\bar{d}\rightarrow W^+ q\bar{q}$, with $q$ of any flavour and $u\bar{d}\rightarrow W^+ gg$. Moreover in
the virtual corrections we have box diagrams which can lead to a different
generation pair in the final state. The divergences from the interference of the $s-$channel
diagrams where a $W$ is emitted from the two initial state legs remain uncancelled, as
in the virtual corrections a $W$ cannot be exchanged between the initial-state quark pair.
The set of diagrams that lead to uncancelled divergences is shown in
Fig.~\ref{udbar} for $u\bar{d}$ scattering. 
\begin{figure}[h]
\centering
\includegraphics[scale=0.7]{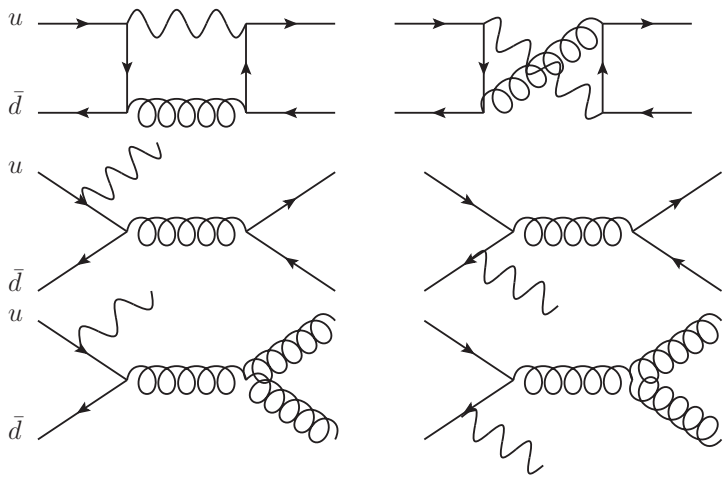} 
\caption{Sample diagrams for initial state $u\bar{d}$ scattering.}
\label{udbar}
\end{figure}
\item \underline{$q\bar{q}\rightarrow Q\bar{Q}$  different generation} In this case the
only LO diagram is an $s-$channel gluon exchange. The virtual corrections
include a diagram where a $W$ boson is exchanged between the $q\bar{q}$ pair, shown in Fig.~\ref{qqbarW} for $u\bar{u}$.
This leads to a DL which cannot be cancelled by the real corrections.
\item \underline{$q\bar{q}\rightarrow q\bar{q}$} The same loop diagram as in the previous case
leads to BN violating DL.
\item \underline{$q\bar{q}\rightarrow Q\bar{Q}$ same generation} In addition to the
vertex correction $s-$channel diagram of the previous two cases, in this case two $t-$channel
box diagrams shown in Fig.~\ref{qqbarbox} are also allowed. These loop diagrams also lead to uncancelled DL.
\begin{figure}[h]
\centering
\includegraphics[scale=0.7]{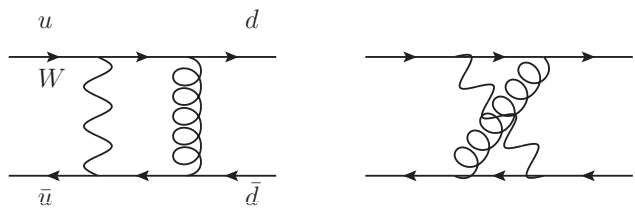} 
\caption{Sample diagrams leading to BN violating logs for initial-state $u\bar{u}$ scattering.}
\label{qqbarbox}
\end{figure}
\end{itemize}

\begin{figure}[h]
\centering
\includegraphics[scale=0.7]{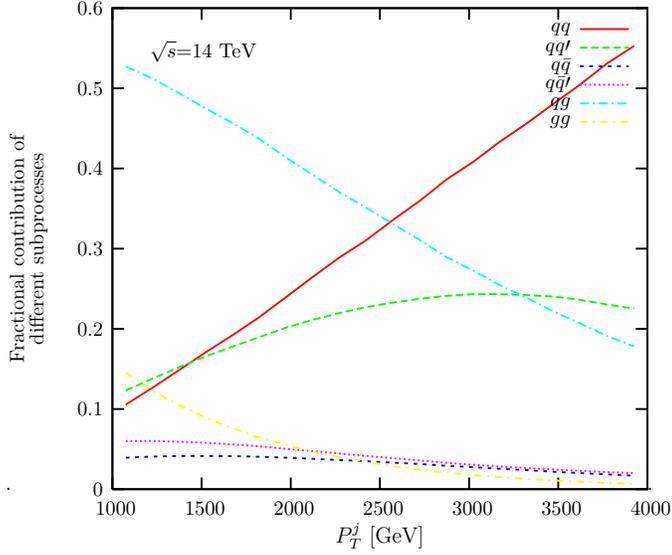} 
\caption{Fractional contributions of different subprocesses as a function of the jet transverse momentum at LO.}
\label{dijet}
\end{figure}

The importance of the BN violating DL can be estimated by considering
Eqs.~(\ref{virtot}) and (\ref{reatot}), and also the relative importance of the
various subprocesses for the dijet cross section at the LHC. This is shown in Fig.~\ref{dijet} with the cross section decomposed into subprocesses based on the initial partonic combination.
A similar decomposition is shown in Table~1 of Ref.~\cite{Moretti:2006ea}, where the contribution to the total EW corrections is decomposed into different subprocesses. 
 To quantify the exact combined effect of real and virtual EW corrections would require a full calculation of both corrections. 

\subsection{Electroweak processes}
\subsubsection{W {\boldmath$+$} Z production}
For comparison with the QCD case, we also consider BN violating effects in a purely electroweak process, i.e. $W +Z$ hadronic production. Once again we will not present any detailed numerical results, rather simply a qualitative discussion. The virtual corrections for this process have been studied in
\cite{Accomando:2001fn} where analytic results obtained using Eq.~(\ref{virtot}) are
presented in the high-energy limit. In this limit longitudinal gauge bosons can
be replaced by the corresponding would-be Goldstone bosons. The virtual corrections are found in \cite{Accomando:2001fn} to decrease the cross section by up to 20\% at $p_T=500$~GeV.
The calculation of the real corrections involves considering the processes $q\bar{q}\rightarrow W^+Z W^-$ and
$q\bar{Q}\rightarrow W^+Z Z$. We have obtained the partonic results for these processes, but the expressions are long and complicated due to the presence of several diagrams. Note that at this NLO-EW order new partonic
PDF combinations are allowed that are not present at leading order (e.g. $u\bar{u}$ scattering does not contribute at leading order but produces the $ W^+ZW^-$ final state at NLO). We therefore expect BN violating logs that
will not cancel even after the addition of the virtual corrections, which are weighted
by a different PDF (LO) combination. 

\subsubsection{W {\boldmath$+$} H production}
The full $\mathcal{O}(\alpha)$ virtual EW and real photon emission corrections to Higgs production in association with a $W$ boson have been studied in \cite{Ciccolini:2003jy}.
Here we are only interested in the DL corrections, which can be extracted and
studied analytically. At LO the process $u\bar{d}\rightarrow W^+H$ has only one Feynman diagram (in the diagonal CKM matrix approximation). The relevant virtual correction diagrams are shown in \cite{Ciccolini:2003jy}. For the real corrections the processes to be considered are
$u\bar{d}\rightarrow W^+H Z$ and the new channel $q\bar{q}\rightarrow W^+H W^-$. For illustration, we consider
 the matrix element squared for the new channel $u(p_1)\bar{u}(p_2)\rightarrow W^+(p_3)H(p_4) W^-$ which factorises in  the high-energy (soft $W^-$) limit:
\begin{eqnarray}\nonumber
|M|^2&=&\frac{\alpha_w}{24\pi c_w^2 s_w^2}\bigg( (3-2s_w^2) \textrm{log}^2\left(\frac{2p_2p_4}{M_W^2}\right)+(3-4s_w^2) 
\textrm{Log}^2\left(\frac{2p_2p_3}{M_W^2}\right)\\
&-&\frac{(9-18s_w^2-8s_w^4)}{6}\textrm{log}^2\left(\frac{2p_3p_4}{M_W^2}\right)\bigg) |M_{\textrm{LO}}|^2,
\end{eqnarray} 
where the intermediate photon contribution is also included. This contribution cannot be cancelled by the virtual corrections, since $u\bar{u}$ scattering does not contribute at leading order.
The corresponding result for $u(p_1)\bar{d}(p_2)\rightarrow W^+(p_3)H(p_4) Z$ in the same limit is 
\begin{eqnarray}\nonumber
 |M|^2&=&\frac{\alpha_w}{72\pi c_w^2 s_w^2}\bigg(-(3-2s_w^2)(3-4s_w^2)\textrm{log}^2\left(\frac{2p_1p_2}{M_Z^2}\right)+3(3-4s_w^2)(1-s_w^2)\textrm{log}^2\left(\frac{2p_1p_3}{M_Z^2}\right)\\ \nonumber
&+&3(3-4s_w^2)\textrm{log}^2\left(\frac{2p_1p_4}{M_Z^2}\right)+3(3-2s_w^2)\textrm{log}^2\left(\frac{2p_2p_3}{M_Z^2}\right)+3(3-2s_w^2)(1-2s_w^2)\textrm{log}^2\left(\frac{2p_2p_4}{M_Z^2}\right)\\
&-&9(1-2s_w^2)\textrm{log}^2\left(\frac{2p_3p_4}{M_Z^2}\right)\bigg)|M_{\textrm{LO}}|^2,
\end{eqnarray}
where we have not used the high-energy approximation $p_3p_4=p_1p_2$ in order to distinguish between logs from the interference of different diagrams. In this form it is 
straightforward to identify how the different prefactors arise from the different $Z$ couplings to $u,d,W$ and Higgs. The DL in this expression exactly cancel if we consider the 1-loop diagrams with virtual $Z$ exchange. 

\section{Conclusions}
We have studied the effect of electroweak corrections on a range of $2\rightarrow 2$ scattering processes at the LHC. 
Existing studies of electroweak corrections generally only take account of  the virtual contributions, as the real emission of EW bosons leads to different final states in the detectors. Nevertheless when sufficiently inclusive measurements are considered, the emission of soft electroweak bosons can significantly counterbalance the effect of large negative Sudakov logs from the 
virtual corrections. Even though the cancellation is never exact because of violations of the Bloch-Nordsieck theorem in electroweak corrections, we have found that the cancellation can be numerically significant. We have only studied the corrections at the DL level, but we can gauge the effect of real radiation, and this has been done at the hadronic level for $\gamma/Z+$jet production, showing that the real radiation effects counterbalance the virtual corrections to a large extent.

We have also studied a sample of other $2\to 2$ QCD and EW processes, identifying potentially large BN violating contributions by considering the relevant Feynman diagrams. For these and other cases, detailed numerical studies going beyond the DL approximation would be needed to quantify the exact impact of the BN violating contributions on hadronic cross sections and these would need
 to include the effect of phase space constraints on the emitted $W,Z$ bosons and their decay products. 

\acknowledgments{E.V. acknowledges financial support from the UK Science and Technology Facilities Council.}

\end{document}